\documentclass[a4paper,10pt]{article}
\usepackage[utf8x]{inputenc}
\usepackage[square, numbers]{natbib}
\usepackage{amsmath}
\usepackage[dvips]{graphics}
\usepackage[dvips]{graphicx}
\usepackage{psfrag}
\usepackage{multirow}
\usepackage{authblk}
\title{Extending HELENA to incompressible plasma rotation parallel to the
magnetic field}
\author[1]{G. Poulipoulis}
\author[1]{ G. N. Throumoulopoulos}
\author[2]{C. Konz} 
\author[ ]{ITM-TF Contributors\footnote{See 
\cite{falchetto2014EurIntTokModITMeffachfirphyres}}}
\affil[1]{Physics Department, University of Ioannina, 451 10 Greece}
\affil[2]{Max-Planck Institut f\"{u}r Plasma Physics, 85748 Garching bei 
M\"{u}nchen, Germany}
\date{}

\begin{document}

\maketitle

\begin{abstract}
Plasma rotation in connection to both zonal and mean (equilibrium) flows can 
play a role in the transitions to the advanced confinement regimes in tokamaks, 
as the L-H transition and the formation of Internal Transport Barriers. For 
incompressible rotation the equilibrium is governed by a generalized  
Grad-Shafranov (GGS) equation and a decoupled  Bernoulli-type equation for the 
pressure. For parallel flow the GGS equation can be transformed  to one 
identical in form with the usual  GS equation. In the present study on the 
basis 
of the latter equation we have extended HELENA, an equilibrium fixed 
boundary solver.  
%integrated in the ITM-TF modeling infrastructure. 
The extended code solves the GGS equation for a variety of the two 
free-surface-function terms involved for arbitrary  Alfv\'en Mach and density 
functions. We have constructed diverted-boundary equilibria pertinent to ITER 
and examined their characteristics, in particular as concerns the impact of 
rotation on certain equilibrium quantities. It turns out that the rotation and 
its shear affect noticeably the pressure and toroidal current density with the  
impact on the current density being stronger  in the parallel direction than  
in 
the toroidal one. Also, the linear stability of the equilibria constructed is 
examined by applying a sufficient condition.
\end{abstract}

%---------------------------
\section{Introduction}

Plasma rotation  affects the equilibrium, stability and transport properties of 
the plasma of a tokamak
device. This has been established
 experimentally as well as  theoretically. 
The appearance of highly peaked density, pressure and temperature profiles, the
 suppression of some instabilities and the creation of transport
barriers, either in the edge region (H-mode) or inside the plasma core (Internal
Transport Barriers), are associated with plasma flow (see for example
\cite{gunter2000SimatthigeleiontemdisinttrabarASDupg}, 
\cite{romanelli2011OveJETres} and the review papers 
\cite{itoh1996rolelefiecon}-\cite{challis2004useinttrabartokpla_2}). The 
transport barriers appear to  be necessary 
ingredient towards the fully non-inductive operation of a future reactor 
\cite{sips2007perimpH-mASDUpgprotoITE}. 
 The flows can be driven externally in connection with electromagnetic power 
and 
neutral beam   injection for  plasma heating and current drive  or can be 
created spontaneously (intrinsic flows).
Though it 
is not fully understood how the flow affects
the confinement properties and especially the formation of transport barriers,
the mode decorrelation and the reduction of turbulence play a significant role. 
Lately there is evidence that the rotation shear is more important
than the flow itself, a fact that re-establishes flow as a key factor for
the future big machines as  ITER and later DEMO in which due to the large
 plasma volume it would be difficult to induce  large flow velocities. 
However, the intrinsic rotation in these machines can be important 
\cite{solomon2007Momconlowtor}. A possible driving mechanism for the creation 
of intrinsic rotation in JET is the pressure gradient
\cite{eriksson1997TorrotICRH-mJET}. 

The MHD equilibria of axisymmetric plasmas,  which can be  starting points of 
stability and transport studies, is governed by the well known  Grad-Shafranov 
(GS) equation. In addition to analytic solutions of this equation, as the 
Solov\'ev solution, a number of fixed and free boundary codes have been 
developed to solve this equation in realistic situations for arbitrary choices 
of the free functions involved including information from experimental data. In 
connection with the present study we refer to the HELENA code, a finite element 
fixed boundary solver of the GS equation in the Integrated Tokamak 
Modelling-Task Force (ITM-TF)  infrastructure. The code is described in Sec. 3. 
In the presence of flow the equilibrium satisfies a generalized  Grad-Shafranov 
(GGS) equation together with a Bernoulli equation involving the pressure 
\cite{morozov1980Steplaflomagfie,hameiri1983equstarotpla}. 
For compressible flows the GGS equation can be either elliptic or hyperbolic 
depending on the value of a  Mach number  associated with the poloidal 
velocity. 
Note that the toroidal velocity is inherently incompressible because of 
axisymmetry. In that case the GGS equation is coupled with the Bernoulli 
equation through the density which is not uniform on magnetic surfaces. A 
number 
of codes have been developed to solve the set of these two equations mainly in 
the first elliptic region which is experimentally accessible \footnote{For a 
typical thermonuclear plasma and magnetic field of the order of 1 Tesla, the 
plasma velocity is of the order of $10$ − $10^2$ Km/sec which lies well within 
the first elliptic region.}, as DIVA \cite{semenzato1984ComsymideMHDfloequ,
strumberger2005NumMHDstastutorrotvisreswalcurhol_2}, FINESSE 
\cite{belien2002FINAxiMHDEquFlo}and FLOW 
\cite{guazzotto2004Numstutokequarbflo}. 
In these codes either an adiabatic or an isothermal equation of state is 
adopted 
associated respectively with either isentropic or isothermal magnetic surfaces. 
For incompressible flow the density becomes a surface quantity and the  GGS 
equation (Eq. (\ref{1}) below) becomes elliptic and  decouples from the 
Bernoulli equation (see Sec. 2). Consequently one has to solve an easier and 
well posed elliptic boundary value problem. In particular for fixed boundaries, 
 convergence to the solution is guaranteed under mild requirements of 
monotonicity for the free functions involved in the GGS equation 
\cite{courant1966Metmatphy}. 
Therefore, it is reasonable to extend existing static equilibrium codes for 
incompressible flows of arbitrary direction. Also, it may be noted that 
uniformity of the density on magnetic surfaces is compatible with the TRANSP 
code \cite{budny1995SimalpparTFTDTsuphigfuspow} which assumes the density as a 
flux function. However, it should be clarified that since density deviations on 
magnetic surfaces have been observed experimentally,  both compressible and 
incompressible equilibrium codes help in obtaining a more complete physical 
understanding.

The stability of  fluids and plasmas in the presence of equilibrium flows non 
parallel to the
magnetic field remains a tough problem reflecting to the lack of necessary and 
sufficient conditions. Only
for parallel  flows few sufficient conditions for linear stability are 
available 
(e.g. 
\cite{friedlander1995stainscrimag}-\cite{morrison2013Stacomredmagequmagins}).  
Also,  stability in the framework of 
normal mode analysis  leads to a complex eigenvalue problem due to the 
antisymmetric properties of the force operator associated with the convective 
flow term in the momentum  equation. Despite  this difficulty, 
normal-mode-based stability studies  have  revealed that 
plasma flow can be a stabilising factor
for the resistive wall mode \cite{garofalo1999Dirobsreswalmodtokitsintplarot, 
garofalo2007StaconreswalmodhigbetlowrotDIIpla, chu2010Staextkinreswalmod, 
igochine2012Phyreswalmod} by mitigating the mode while in turn the mode is 
braking the 
rotation. The rotation threshold, under which the mode is observed, appears to 
be a small fraction of the Alfv\'{e}n frequency, though the experimental 
results on that point are not clear 
\cite{garofalo2007StaconreswalmodhigbetlowrotDIIpla}. In 
addition, a connection between the plasma rotation and the error field is 
believed to affect the stabilization of the mode. 
Regarding the neoclassical tearing modes, reducing plasma rotation results in a 
deterioration of  stability and an increase in the size of saturated islands 
\cite{politzer2008InftorrottrastahybsceplaDII,
lahaye2010Islstreffplafloteasta}. Therefore plasma rotation seems to be a key 
element for  most -if not all- of the Advanced Tokamak Scenarios. 
%Understanding the way plasma rotation affects the equilibrium properties of a 
%tokamak is of importance, especially in preparation towards ITER operation. 

The aim of the present study is to extended the well known and widely used
%for the Integrated Tokamak Modelling - Task Force (ITM-TF) purposes, 
code
HELENA\cite{konz2011FirphyappIntTokModITMtootoMHDstaanaexpdatITEsce} for 
incompressible rotation parallel to the magnetic field  and to evaluate the 
impact of  rotation on the equilibrium characteristics. It is noted that  
poloidal equilibrium flow components are included in the ITM-TF for the first 
time.  The 
extension of the code is based on Eq. (\ref{1}) which under an integral 
transformation can be but in a form identical with the usual GS equation. In 
addition we will examine the linear stability of the equilibria constructed 
 by means of a pertinent sufficient condition 
\cite{throumoulopoulos2007sufconlinstamagequfiealiincflo_2}. 

The report is organised as follows. The GGS equation for plasmas with 
incompressible flow is reviewed in Sec. 2. In Sec. 3 the HELENA code is 
extended for rotation parallel to the magnetic field, particular equilibria are 
constructed and the impact of rotation and its shear on certain equilibrium 
quantities is examined. The linear stability of the equilibria constructed is 
studied  by applying the stability condition of Ref. 
\cite{throumoulopoulos2007sufconlinstamagequfiealiincflo_2} in Sec. 4 and the 
results are compared with those of previous studies mainly based on analytic 
equilibrium solutions. Section 5 summarises the main conclusions.

%-----------------------------------------------------------------------
\section{Generalized Grad Shafranov equation}

The equilibrium of a magnetically confined  plasma with incompressible flow of 
arbitrary direction satisfies the following generalized  GS  equation 
(GGS) \cite{tasso1998Axiidemagequincflo_2, 
poulipoulis2005Torflochamagtopequeig}: 
\begin{align}
 (1-M_p^2) \Delta^\star \psi -
         \frac{1}{2}(M_p^2)^\prime |\nabla \psi|^2
                     + \frac{1}{2}\left(\frac{X^2}{1-M_p^2}\right)^\prime 
\nonumber  \\
+\mu_0 R^2 P_s^\prime + \mu_0 \frac{R^4}{2}\left[ 
\frac{\rho(\Phi^\prime)^2}{1-M_p^2}\right]^\prime
    = 0
 \label{1}
 \end{align}
 Here,  the poloidal magnetic flux function $\psi(R,z)$  labels the magnetic 
surfaces, where  ($R,\phi, z$) are cylindrical coordinates with $z$ 
corresponding to the axis of symmetry; $M_p(\psi)$ is
 the   Mach function of the poloidal fluid velocity with respect to the
poloidal  Alfv\'en velocity; $X(\psi)$ relates to the toroidal magnetic
 field, $B_\phi=I/R$,  through $I=X/(1-M_p^{2})$; $\Phi(\psi)$ is the 
electrostatic potential; for vanishing flow the surface function $P_s(\psi)$
  coincides with the pressure; $B$ is the magnetic field modulus
  which can be expressed in terms of surface functions and $R$;  
$\Delta^\star=R^2\nabla\cdot(\nabla/R^2)$;
  and the prime denotes derivatives  with respect to $\psi$.
  Because of incompressibility the density $\rho(\psi)$ is also a surface 
quantity and the Bernoulli equation for the pressure decouples from (\ref{1}):
\begin{equation}
 P=P_s(\psi) - \varrho \left( \frac{\upsilon^2}{2} - \frac{R^2
 (\Phi^\prime)^2}{1-M_p^2}\right)
                          \label{2}
 \end{equation}
 where $\upsilon$ is the velocity modulus. The  quantities
$M_p(\psi)$,  $X(\psi)$,
$P_s(\psi)$, $\rho(\psi)$ and $\Phi(\psi)$ are free functions.
Derivation of Eq. (\ref{1}) is based on the following two steps: first express 
the divergence free fields ($\bf B$, $\bf j$ and $\rho {\bf v}$) in terms of 
scalar quantities and second, project the momentum equation,
$\rho({\bf v}\cdot\nabla){\bf v}={\bf j}\times{\bf B}-\nabla P$,
and Ohm's law, along the toroidal direction, ${\bf B}$ and $\nabla\psi$. The 
projections yield four first integrals in the form of surface quantities
and  Eqs. (\ref{1}) and \ref{2}.
Details are given   in  
\cite{tasso1998Axiidemagequincflo_2}-\cite{throumoulopoulos2008Sidaxiequincflo}
. 
As already mentioned in Sec. 1 the decoupling of the GS equation from the 
pressure equation due 
to the incompressibility is the major difference between the current work and 
other models considering compressible flows associated with  alternative 
equations of state. In the latter cases, the 
pressure equation needs to be solved simultaneously with the GS 
equation, while in our case, one has to just solve the GS under appropriate 
boundary conditions.
Then, the Bernoulli equation is  used as a formula to obtain the 
pressure.

Eq. (\ref{1}) can be  simplified  by  the  transformation
\begin{equation}
u(\psi) = \int_{0}^{\psi}\left\lbrack 1 -
M_p^{2}(f)\right\rbrack^{1/2} df
                                            \label{3}
\end{equation}
under which  (\ref{1})  becomes
\begin{align}
  \Delta^\star u
+ \frac{1}{2}\frac{d}{du}\left(\frac{X^2}{1-M_p^2}\right)  +
\mu_0 R^2\frac{d P_s}{d u} \\
+ \mu_0\frac{R^4}{2}\frac{d}{du}\left[\rho\left(\frac{d 
\Phi}{du}\right)^2\right] = 0
                            \label{4}
\end{align}
 Note  that no quadratic term as $|{\bf\nabla}u|^{2}$ appears
any more in (\ref{4}). It is emphasized that once a solution of (\ref{4}) 
is obtained,  the equilibrium can be completely  constructed with calculations 
in the $u$-space by employing (\ref{3}), and  the inverse transformation
\begin{equation}
\psi(u) = \int_{0 }^{u}\left\lbrack 1 -
M_p^{2}(f)\right\rbrack^{-1/2} df
                                            \label{5}
\end{equation}
Specifically one finds  
\begin{align}
P=P_s(u)-\varrho(u)\left[\frac{\upsilon^2}{2}-R^2\left(\frac{d\Phi(u)}{du}
\right)^2\right] \label{pres1} \\
\vec{B}=I(\psi)\vec{\nabla}\phi -\vec{\nabla}\phi\times\vec{\nabla}\psi= 
I(u)\vec{\nabla}\phi-\frac{d\psi}{du}\vec{\nabla}\phi\times\vec{\nabla}u 
\label{magfiel} \\
\vec{J}=\frac{1}{\mu_0}\left(-\Delta^*\psi\vec{\nabla}\phi+\vec{
\nabla}\phi\times \vec{\nabla}I(\psi)\right)= \nonumber \\
\frac{d\psi}{du}R^2\vec{\nabla}\left(\frac{\vec{\nabla}u}{R^2}\right)+ 
\vec{\nabla} u \cdot \vec{\nabla}\frac{d\psi}{du} +\frac{dI(u)}{du}\vec{ 
\nabla}\phi\times \vec{\nabla}u \label{curden}\\
\vec{E}=-\vec{\nabla} \Phi= -\frac{d \Phi(\psi)}{d \psi} \vec{\nabla} \psi= 
-\frac{d \Phi(u)}{d u} \vec{\nabla} u
\end{align}
For parallel flows ($\Phi^\prime=0$),  Eq.  (\ref{4}) reduces in form to  the 
usual GS equation. Also, in this case  it can be shown
that the poloidal, toroidal and total velocity Alfv\'{e}n Mach numbers 
are exactly equal; henceforth the total velocity Mach number will be indicated 
by $M$.  Moreover, in this case, setting
\begin{equation}
\varrho\vec{\upsilon}=K\vec{B} \label{veloc}
\end{equation}
where $K$ is a scalar function, applying the divergence operator and taking  
into account the continuity equation, 
$\vec{\nabla}\cdot\left(\varrho\vec{\upsilon}\right)=0$, one obtains 
$\vec{\nabla}K\cdot\vec{B}=0 $ which means that the free function K is a 
surface 
quantity \cite{throumoulopoulos2003axiresmagequflofrePfidif}
$$
K=K(\psi)
$$
Making use of Eq. (\ref{veloc}) the Bernoulli equation \ref{2}, in the case of 
parallel plasma rotation can be written as
\begin{equation}
 P=P_s(\psi) - \frac{1}{2\mu_0}M^2B^2(\psi, R)= P_s(u) 
- \frac{1}{2\mu_0}M^2B^2(u, R)  \label{pres2}
\end{equation}

%-----------------------------------------------------------------------
\section{Extension of the HELENA code}

The code HELENA, is a fixed boundary equilibrium solver 
\cite{huysmans1991IsobicHerelesolGraequ} available on
the EFDA ITM Gateway and used for ITM-TF purposes. The static  GS
equation used in the code is written as: 
\begin{equation}
\Delta^*\psi = -F\frac{dF}{d\psi} - \mu_0R^2\frac{dP}{d\psi} = -\mu_0Rj_{tor}
\label{eq5}
\end{equation}
HELENA solves Eq. (\ref{5}) for a
toroidal axisymmetric plasma by use of isoparametric bicubic Hermite finite 
elements. The main ingredients  of the numerical algorithm is the Galerkin 
method, a non linear iteration scheme leading to a set of linear equations for 
each step of the iteration, and the use of straight field line coordinates. The 
function $\psi$   is specified to assume a certain
value  on a predefined boundary which for HELENA is the last closed flux 
surface 
of its computational domain.
This numerical technique yields very accurate solutions and has good 
convergence 
properties. Details are provided in \cite{KonzHELENA}.

Comparison of 
(\ref{4}) for parallel plasma rotation ($\Phi^\prime=0$)  with  (\ref{eq5}) 
implies the following correspondence: 
\begin{align}
\psi \longleftrightarrow u 
\label{eq6} \\
F\frac{dF}{d\psi} \longleftrightarrow \frac{1}{2}\frac{d}{du}\left( \frac{X^2}{1
- M^2}\right) \label{eq7} \\
P(\psi) \longleftrightarrow P_s(u) 
\label{eq8}
\end{align}
Therefore, the solver of the static code HELENA can be used to calculate the 
stationary equilibrium for parallel plasma rotation, 
though the output will no longer correspond to the ``natural'' quantities in the
$\psi$-space. In order to preserve compatibility with the conventions 
established in the former EFDA ITM-TF
the calculated by the solver quantities (now in the u-space) must be
mapped to the ``natural'' $\psi$-space. For the mapping one must consider the
following correspondence for the basic quantities:

\begin{align}
P_{\mbox{\scriptsize HELENA}}\longleftrightarrow P_s \label{eq8a} \\
F_{\mbox{\scriptsize HELENA}}\longleftrightarrow \frac{X}{\sqrt{1-M^2}} \\
\psi_{\mbox{\scriptsize HELENA}}\longleftrightarrow u \label{eq8c}
\end{align}
By using Eqs. (\ref{magfiel}), (\ref{curden}) for $\Phi^\prime=0$, Eq. 
(\ref{pres2}) and (\ref{eq8a})-(\ref{eq8c}), we get the following expressions 
for the magnetic field, current density and pressure. 
\begin{align}
\addtolength{\itemsep}{-5mm}
\vec{B}=\frac{F_{\mbox{\scriptsize
HELENA}}}{\sqrt{1-M^2}}\vec{\nabla}\phi-\frac{1}{\sqrt{1-M^2}}
\vec{\nabla}\phi\times\vec{\nabla}u \label{eq:9} \\
\vec{J}=\left[\frac{-1}{\sqrt{1-M^2}}\Delta^*u+\frac{1}{2}\frac{1}{(1-M^2)^{3/2}
} \frac{dM^2}{du}
|\vec{\nabla}u|^2\right]\vec{\nabla}\phi+  \nonumber \\
\frac{d}{du}\left(\frac{F_{\mbox{
\scriptsize HELENA}}}{
\sqrt{1-M^2}}\right) \vec{\nabla} \phi\times\vec{\nabla}u \label{eq:10} \\
 P=P_{\mbox{\scriptsize HELENA}}-\frac{1}{2 \mu_0
R^2}\frac{M^2}{1-M^2}\left(F_{\mbox{\scriptsize HELENA}}^2+|\vec{\nabla}
u|^2\right) 
\label{eq:11}
\end{align}
where the subscript HELENA refers to the computed by the solver quantities.

\begin{figure}[ht!]
\begin{center}
\psfrag{z}{$z(m)$}
\psfrag{R}{$R(m)$}
\includegraphics[scale=0.45]{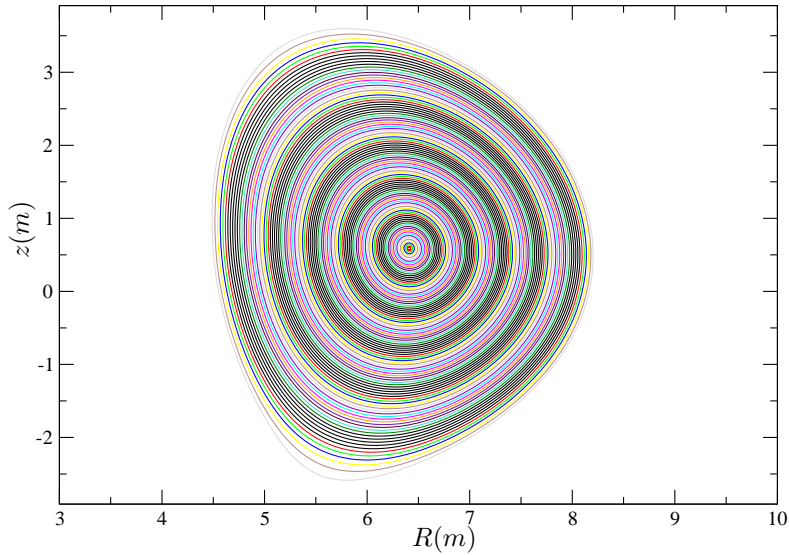}
\caption{The magnetic surfaces of a paramagnetic  equilibrium associated with 
the values of Table 1. The magnetic axis is located 
at ($R_{a}=6.4099$ m, $z_{a}=0.588$ m)  where the toroidal magnetic field is 
5.77 T with respective vacuum value 5.178 T. The integral transformation 
(\ref{3}) leaves the magnetic surfaces intact; therefore this configuration is 
unaffected by rotation as long as the input to the code is kept fixed.}
\label{fig:surf}
\end{center}
\end{figure}

It is emphasized that any solution of (\ref{eq5}),  and therefore of the 
extended code, holds for arbitrary Mach numbers $M(u)$ and densities 
$\varrho(u)$. We have obtained a variety of equilibria by running the extended 
code. An  example showing the magnetic surfaces of an ITER-relevant solution  
associated with input values of Table \ref{tab:0}  is given in Fig.  
\ref{fig:surf}.
\begin{table}
\centering
\caption{Values of some of the basic quantities of shot 35441.}
\begin{tabular}{ c c c c c c c}
\hline \hline
$R_0$ & $B_{\phi 0}$ & $I_{plasma}$ & $q_{min}$ & $\beta_t$ & $\psi_{axis}$ & 
$\psi_{bound}$ \\
6.34 m  & 5.178 T & 14.46 MA & 0.86 & 0.0103 & 0.0 Wb & 79.09 Wb  \\
 \hline
 \end{tabular}
 \label{tab:0}
\end{table}
This input for the solver is the 35441 shot of the ITM database 
within the infrastructure of the former EFDA ITM-TF. The input parameters of 
the 
code remained the same and together with the boundary, the pair of the free 
flux 
functions $P'$ and $FF'$ were chosen as input along the geometric centre, the 
vacuum magnetic field and $\psi$ at the boundary. The results were obtained by 
means of a simple Kepler workflow which reads the input CPO of the shot, runs 
the HELENA code and stores the results in a CPO saved in the user's local 
database. It must be noted here that all the quantities that are stored in the 
CPO are in the $\psi$-space either via the transformation (such as the current 
densities, the pressure etc. as mentioned before) or by directly applying the 
inverse transformation (\ref{5}), such as the poloidal flux function $\psi$. 

\begin{figure}[ht!]
\begin{center}
\psfrag{mach}{$M$}
\psfrag{psi}{$\psi(Wb)$}
\psfrag{stationaryiter20323 }[c][c]{$M_0=0.03$, $m=2$, 
$n=3$\hspace{1cm}}
\psfrag{stationaryiter20344 }[c][c]{$M_0=0.03$, $m=4$, 
$n=4$\hspace{1cm}}
\psfrag{stationaryiter20623 }[c][c]{$M_0=0.06$, $m=2$, 
$n=3$\hspace{1cm}}
\includegraphics[scale=0.45, angle=270]{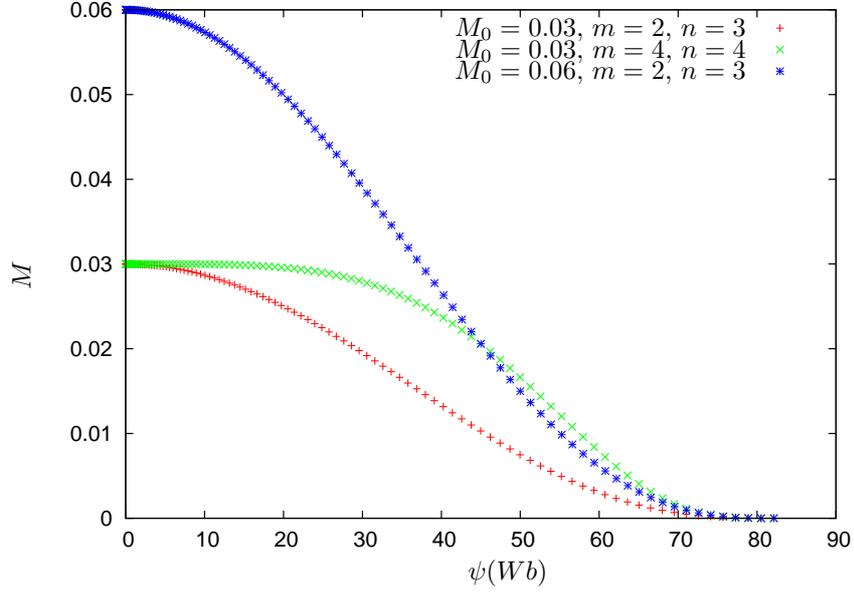}
\caption{Plots of the on-axis peaked Mach number profile (Eq. (\ref{prof1})) 
with respect to 
$\psi$  for various values of the profile parameters.}
\label{fig:mach1}
\end{center}
\end{figure}

\begin{figure}[ht!]
\begin{center}
\psfrag{mach}{$M$}
\psfrag{psi}{$\psi(Wb)$}
\psfrag{stationaryiter30224 }[c][c]{$M_0=0.02$, $m=2$, 
$n=4$\hspace{1cm}}
\psfrag{stationaryiter30552 }[c][c]{$M_0=0.05$, $m=5$, 
$n=2$\hspace{1cm}}
\psfrag{stationaryiter30352 }[c][c]{$M_0=0.03$, $m=5$, 
$n=2$\hspace{1cm}}
\includegraphics[scale=0.45, angle=270]{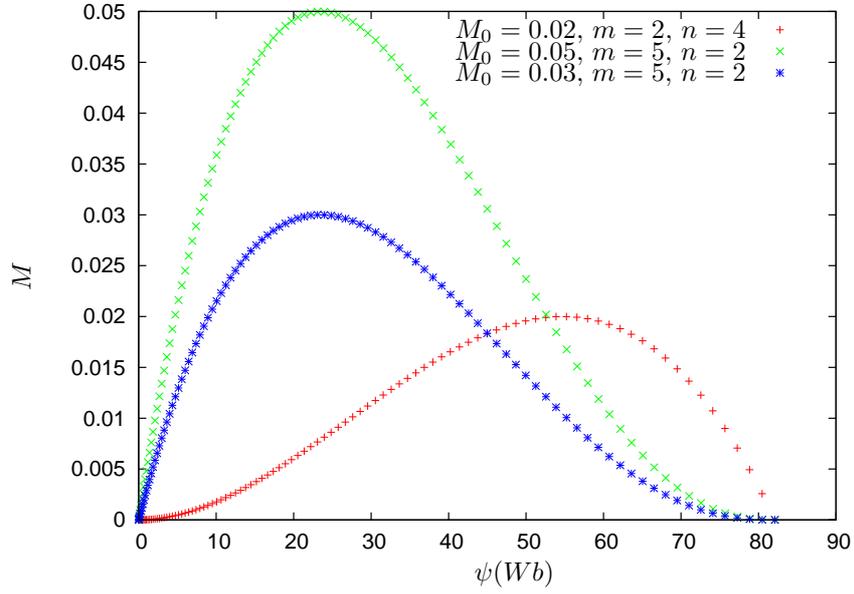}
\caption{Plots of ITER-relevant off-axis peaked Mach number profiles (Eq. 
(\ref{prof2})) with respect to 
$\psi$  for various values of the free parameters. }
\label{fig:mach2}
\end{center}
\end{figure}

To construct completely particular equilibria  we have
made several choices of the free function $M^2(u)$, e.g., 
\begin{eqnarray}
M^2& =& M_0^2 \left(u^m - u_b^m\right)^n
\label{prof1} \\
M^2&=& C \left[\left(\frac{u}{ u_b}\right)^m\left( 
1-\left(\frac{u}{ u_b}\right)\right)\right]^n
\label{prof2} 
\end{eqnarray}
 with
$$
C=M_0^2\left(\frac{m+n}{m}\right)^m \left(\frac{n}{m+n}\right)^n
$$
Here, $ u_b$ refers to the plasma boundary;  the free parameter
$M_0^2$ correspond to the maximum value of $M^2$; and $m$ and 
$n$ are related to flow shear and the position of the maximum $M^2$. In 
particular, (\ref{prof1}) is  peaked on- while (\ref{prof2}) peaked off-axis in 
connection with respective auxiliary heating of tokamaks. Note that for 
parallel 
rotation there is no need to specify the  density because it can be eliminated. 
Plots of $M$ in connection to (\ref{prof1}) and (\ref{prof2})  for different 
values of the free parameters are given in Figs. 
\ref{fig:mach1} and \ref{fig:mach2} respectively.
In general the rotation  has a rather weak contribution to Eq.
(\ref{4}).  However, as already mentioned in Sec. 1,   it is the velocity
shear that is more important for the transition to improved confinement
regimes in tokamaks than the velocity amplitude. 

\begin{figure}[ht!]
\begin{center}
\psfrag{pres}{$P(Pa)$}
\psfrag{r}{$R(m)$}
\psfrag{analyticp }[c][c]{Analytical\hspace{0.8cm}}
\psfrag{numericalp }[c][c]{Numerical\hspace{0.55cm}}
\includegraphics[scale=0.45, angle=270]{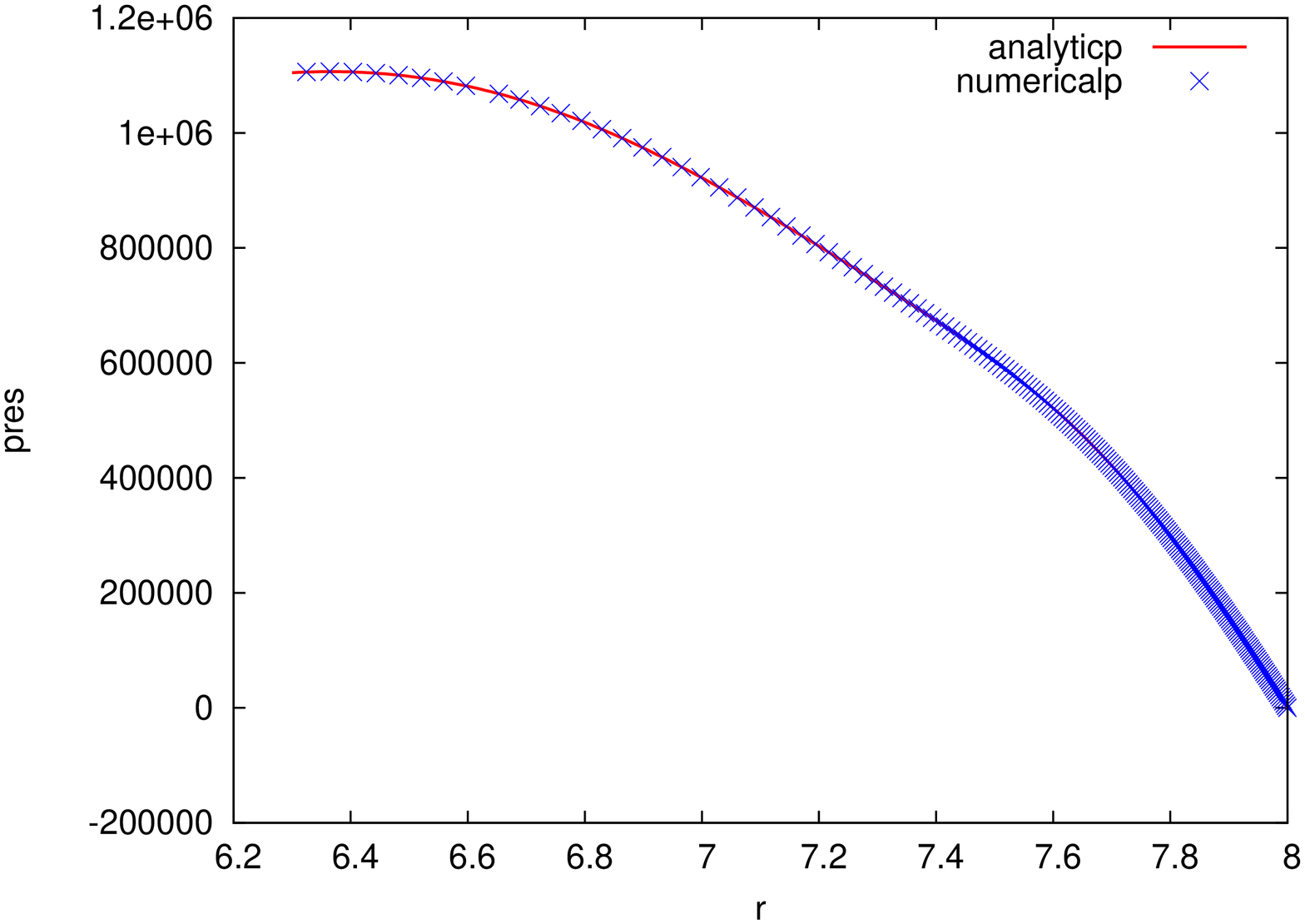}
\caption{Plots of the pressure with respect to the radial distance from the 
axis 
of symmetry  at the midplane $z=0$ for the generalized Solov\'ev solution of 
Ref. \cite{simintzis2001Anamagequmagconplasheflo} and the numerical one for 
the peaked on-axis Mach number profile (\ref{prof1}) with $M_0=0.1$, $n=2$, 
$m=3$.}
\label{fig:prs_sol}
\end{center}
\end{figure}

\begin{figure}[ht!]
\begin{center}
\psfrag{jphi}{$J_{\phi}(A/m^2)$}
\psfrag{r}{$R(m)$}
\psfrag{analyticjphi }[c][c]{Analytical\hspace{0.8cm}}
\psfrag{numericaljphi }[c][c]{Numerical\hspace{0.52cm}}
\includegraphics[scale=0.45, angle=270]{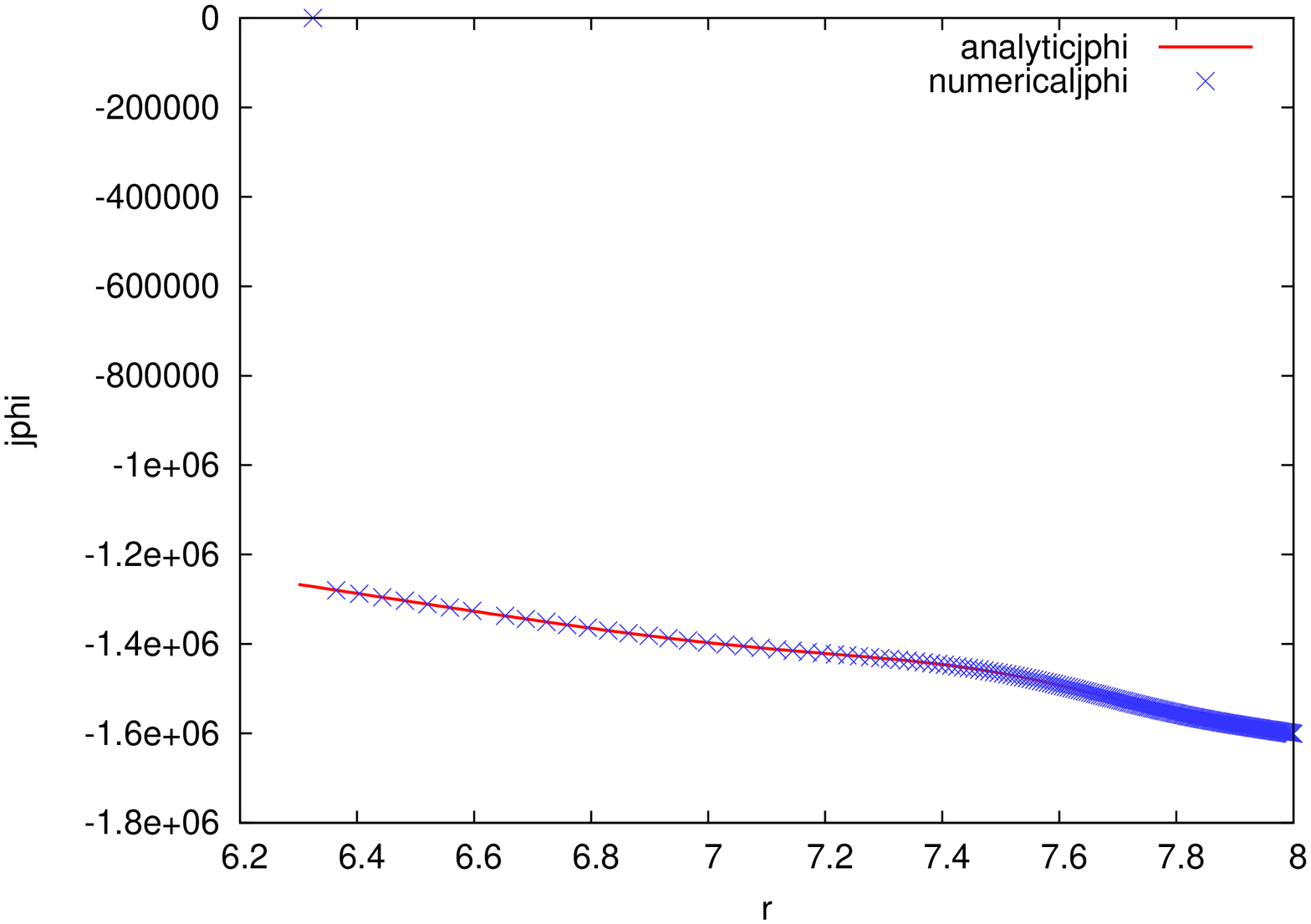}
\caption{Plots of the toroidal current density   with respect to the radial 
distance from the axis of symmetry  at the midplane $z=0$ 
for the generalized  Solov\'ev solution of Ref. 
\cite{simintzis2001Anamagequmagconplasheflo} and the numerical one for 
the peaked on-axis Mach number profile (\ref{prof1}) with $M_0=0.1$, $n=2$, 
$m=3$.}
\label{fig:jphi_sol}
\end{center}
\end{figure}

It is noted here that although Eq. (\ref{4}) (with $\Phi^\prime=0$) is 
identical 
in form with (\ref{eq5}) we benchmarked the extended code against a generalized 
 Solov\'ev solution for parallel  rotation involving $M^2$ 
\cite{simintzis2001Anamagequmagconplasheflo}. The agreement is very good  as 
shown in Figs. \ref{fig:prs_sol} and \ref{fig:jphi_sol} for the pressure and 
the toroidal current density respectively.

By varying the rotation parameters ($M_0$, $n$, $m$) of  the Mach number 
profile we examined the effect of  rotation in certain  of the basic 
equilibrium quantities. It is noted here that  we  consider the system without 
external momentum and energy sources and therefore the total energy of the 
system is kept constant  regardless of the the nature of the rotation.  Also,  
in the case of intrinsic rotation  sources are not necessary. This is consistent 
with  a desirable long term operation of a tokamak reactor after potential 
initial external energy and momentum sources have been removed. 

\begin{figure}[ht!]
\begin{center}
\psfrag{p}{$P(Pa)$}
\psfrag{r}{$R(m)$}
\psfrag{staticcase2m00m0n0 }[c][c]{Static Case, $M_0=0$\hspace{0cm}}
\psfrag{stationaryiter30224 }[c][c]{$M_0=0.02$, $m=2$, 
$n=4$\hspace{1cm}}
\psfrag{stationaryiter30552 }[c][c]{$M_0=0.05$, $m=5$, 
$n=2$\hspace{1cm}}
\includegraphics[scale=0.45, angle=270]{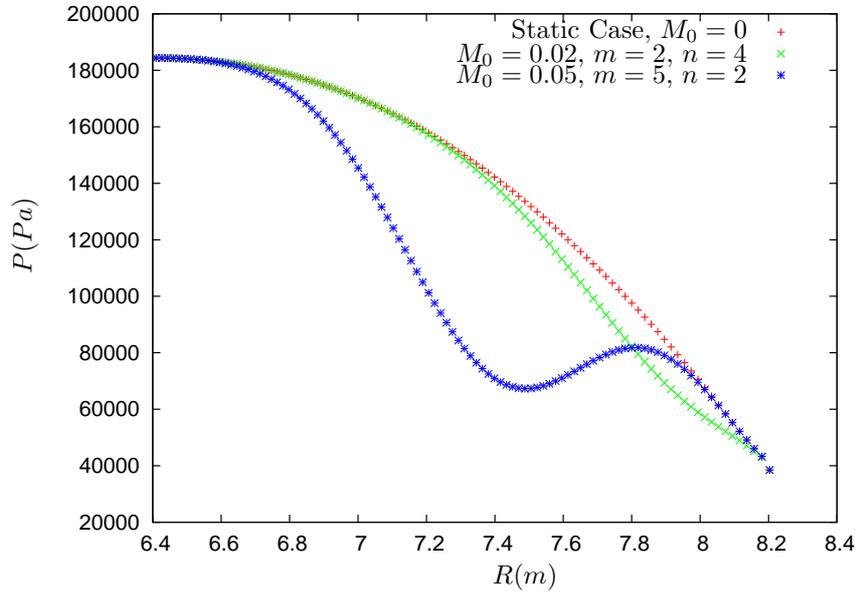}
\caption{Plots of the pressure with respect to the radial distance from the 
axis 
of symmetry at the midplane $z=0$ for various Mach number profiles peaked 
off-axis.}
\label{fig:prs1}
\end{center}
\end{figure}

\begin{figure}[ht!]
\begin{center}
\psfrag{p}{$P(Pa)$}
\psfrag{r}{$R(m)$}
\psfrag{staticcase2m00m0n0 }[c][c]{Static Case, $M_0=0$\hspace{0cm}}
\psfrag{stationaryiter20323 }[c][c]{$M_0=0.03$, $m=2$, 
$n=3$\hspace{1cm}}
\psfrag{stationaryiter20344 }[c][c]{$M_0=0.03$, $m=4$, 
$n=4$\hspace{1cm}}
\psfrag{stationaryiter20623 }[c][c]{$M_0=0.06$, $m=2$, 
$n=3$\hspace{1cm}}
\includegraphics[scale=0.45, angle=270]{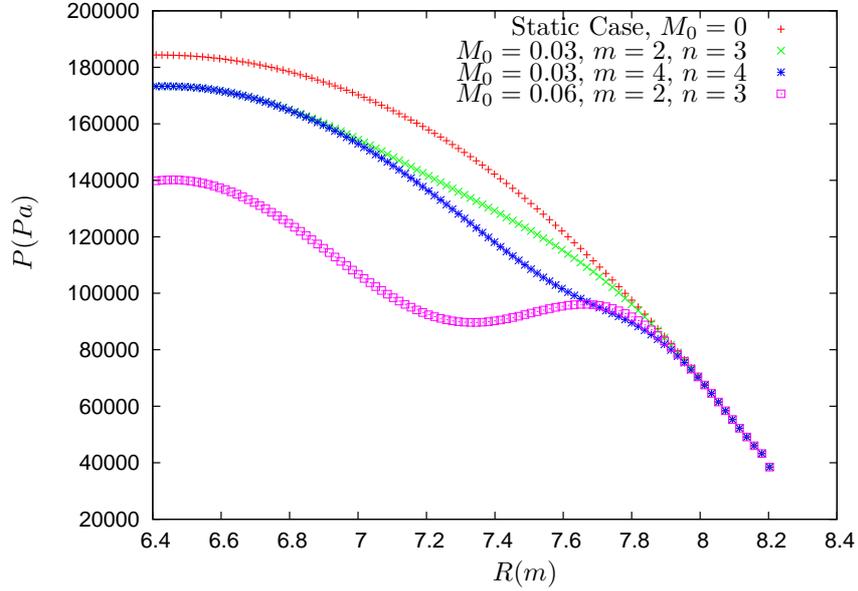}
\caption{Plots of the pressure with respect to the radial distance from the 
axis 
of symmetry on  the midplane $z=0$ for various  Mach number profiles peaked 
on-axis.}
\label{fig:prs2}
\end{center}
\end{figure}

Regarding the pressure,  on the basis of Eq. (\ref{eq:11}) for given $P_s$ 
plasma rotation is 
expected to reduce the pressure values in comparison with the static ones. It 
appears that 
both the values of the Mach number and the shear of its profile affect the 
pressure. This result is different than that from code DIVA 
\cite{strumberger2005NumMHDstastutorrotvisreswalcurhol_2} due either to 
the direction of the rotation or to compressibility.
It is worth noting that in the case of the off-axis localised 
rotation (Fig. \ref{fig:prs1}), the 
effect on the pressure profile is significant compared to the on-axis case 
(Fig. 
\ref{fig:prs2}) for the same values of $M_0$, $n$ and $m$. This is reasonable 
since in the off-axis case the values of the Mach 
number profile in Eq. (\ref{eq:11}) away from the magnetic axis are comparably 
larger than  the 
on-axis-profile ones with respect to the static pressure term. Though for 
experimentally accessible  velocity values on axis the effect on the pressure 
profile is small, the change in the shape of the profile can possibly affect 
the 
growth rate of pressure gradient driven modes as well as the intrinsic rotation 
\cite{eriksson1997TorrotICRH-mJET}. This also holds for the  off-axis rotation; 
also the values of the $M_0$ that affect significantly the 
pressure are possibly accessible experimentally. In addition this type of 
profile is more probable for ITER which will exhibit a larger moment of inertia 
compared to present day tokamaks \cite{devries2008ScarotmomconJETpla}. One 
more notable point is that for  off-axis  rotation the pressure 
profile is compatible with experimental results associated with the formation 
of either Internal Transport Barriers  or edge barriers (H-mode). Specifically 
the pressure profile exhibits a steepness region where the 
maximum of the Mach number profile is located. In fact  plasma rotation is 
associated with mode decorrelation resulting in  the reduction of  transport 
coefficients and eventually the formation of a transport barrier. On physical 
grounds, the increase of the steepness of the pressure profile is related  with 
the formation of this barrier in connection with the flow.
Also,  since the rotation decreases the pressure and increases 
the toroidal magnetic field (see Eqs. (\ref{pres1}) and (\ref{eq:9})) it is 
expected that in the presence of rotation the toroidal beta will decrease. Both 
the maximum value of $M^2$ and its shear contribute to this reduction as 
indicated by the results quoted in Table \ref{tab:1}.

\begin{table}
\centering
\caption{Values of the $\beta_{tor}$ for various rotation cases.}
\begin{tabular}{| c | c | c | c | c |}
\hline
 Static & \multicolumn{2}{c}{On-axis} \vline  &\multicolumn{2}{c}{Off-axis} 
\vline  \\
 \hline 
 \mbox{} & $M_0=0.03$ & $M_0=0.03$ & $M_0=0.05$ & $M_0=0.02$ \\
\mbox{} & $m=3$ $n=3$ & $m=4$ $n=4$  & $m=5$ $n=2$ & $m=2$ $n=4$ \\
 \hline
 0.0100 & 0.0091 & 0.0089 & 0.0075 & 0.0093  \\
 \hline
 \end{tabular}
 \label{tab:1}
\end{table}

\begin{figure}[ht!]
\begin{center}
\psfrag{jphi}{$J_\phi(A/m^2)$}
\psfrag{r}{$R(m)$}
\psfrag{staticcaseiter2000 }[c][c]{Static Case, $M_0=0$\hspace{0.4cm}}
\psfrag{stationaryiter20323 }[c][c]{$M_0=0.03$, $m=2$, 
$n=3$\hspace{1cm}}
\psfrag{stationaryiter20344 }[c][c]{$M_0=0.03$, $m=4$, 
$n=4$\hspace{1cm}}
\psfrag{stationaryiter20623 }[c][c]{$M_0=0.06$, $m=2$, 
$n=3$\hspace{1cm}}
\includegraphics[scale=0.45, angle=270]{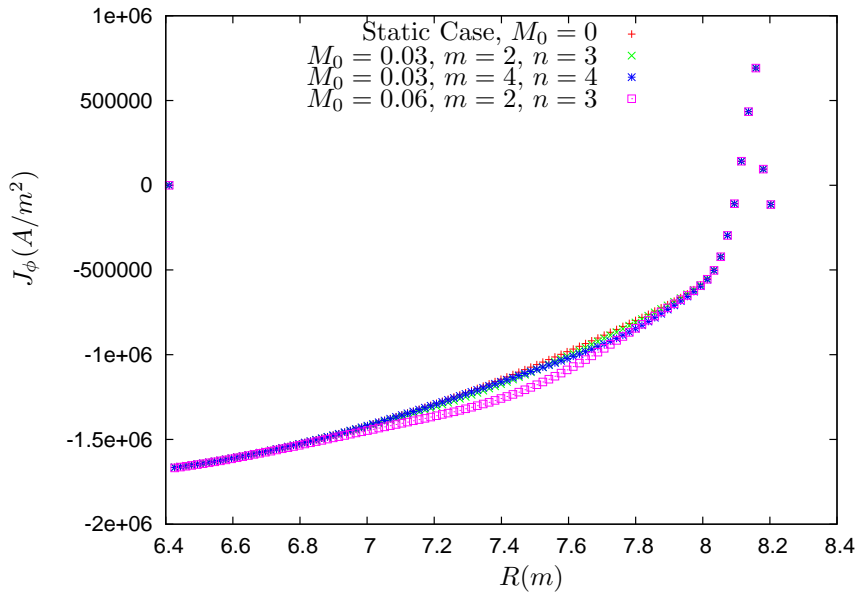}
\caption{Plots of the toroidal current density versus the radial distance from 
the axis of symmetry  on 
the midplane $z=0$ for various  Mach number profiles peaked on-axis.}
\label{fig:jphi1}
\end{center}
\end{figure}

\begin{figure}[ht!]
\begin{center}
\psfrag{jphi}{$J_\phi(A/m^2)$}
\psfrag{r}{$R(m)$}
\psfrag{staticcaseiter2000 }[c][c]{Static Case, $M_0=0$\hspace{0.4cm}}
\psfrag{stationaryiter30224 }[c][c]{$M_0=0.02$, $m=2$, 
$n=4$\hspace{1cm}}
\psfrag{stationaryiter30552 }[c][c]{$M_0=0.05$, $m=5$, 
$n=2$\hspace{1cm}}
\includegraphics[scale=0.45, angle=270]{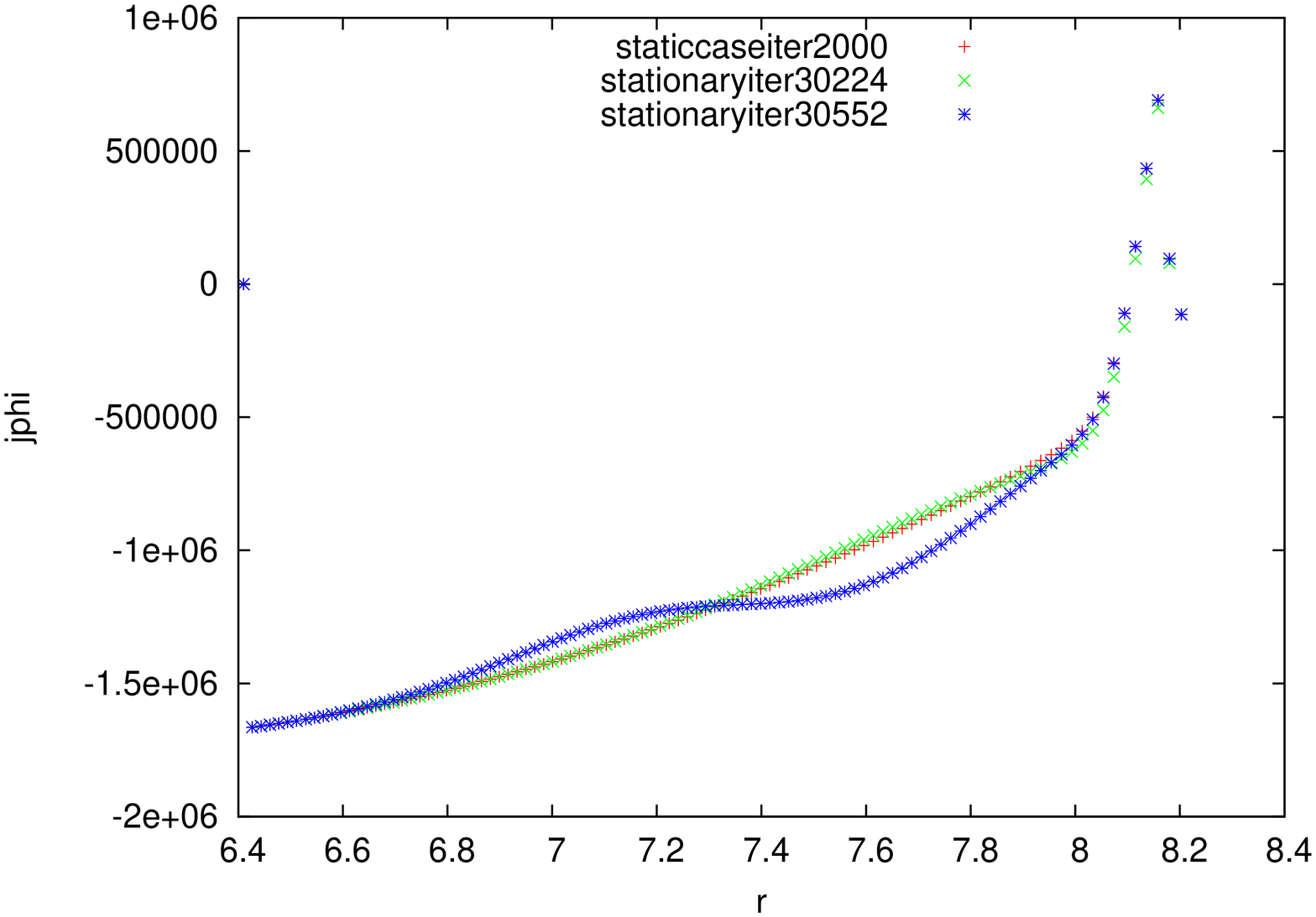}
\caption{Plots of the toroidal current density versus the radial distance from 
the axis of symmetry on the midplane $z=0$ for various  Mach number profiles 
peaked off-axis.}
\label{fig:jphi2}
\end{center}
\end{figure}

\begin{figure}[ht!]
\begin{center}
\psfrag{jpar}{$J_\|(A/m^2)$}
\psfrag{r}{$R(m)$}
\psfrag{staticcaseiter2000 }[c][c]{Static Case, $M_0=0$\hspace{0.4cm}}
\psfrag{stationaryiter20323 }[c][c]{$M_0=0.03$, $m=2$, 
$n=3$\hspace{1cm}}
\psfrag{stationaryiter20344 }[c][c]{$M_0=0.03$, $m=4$, 
$n=4$\hspace{1cm}}
\psfrag{stationaryiter20623 }[c][c]{$M_0=0.06$, $m=2$, 
$n=3$\hspace{1cm}}
\includegraphics[scale=0.45, angle=270]{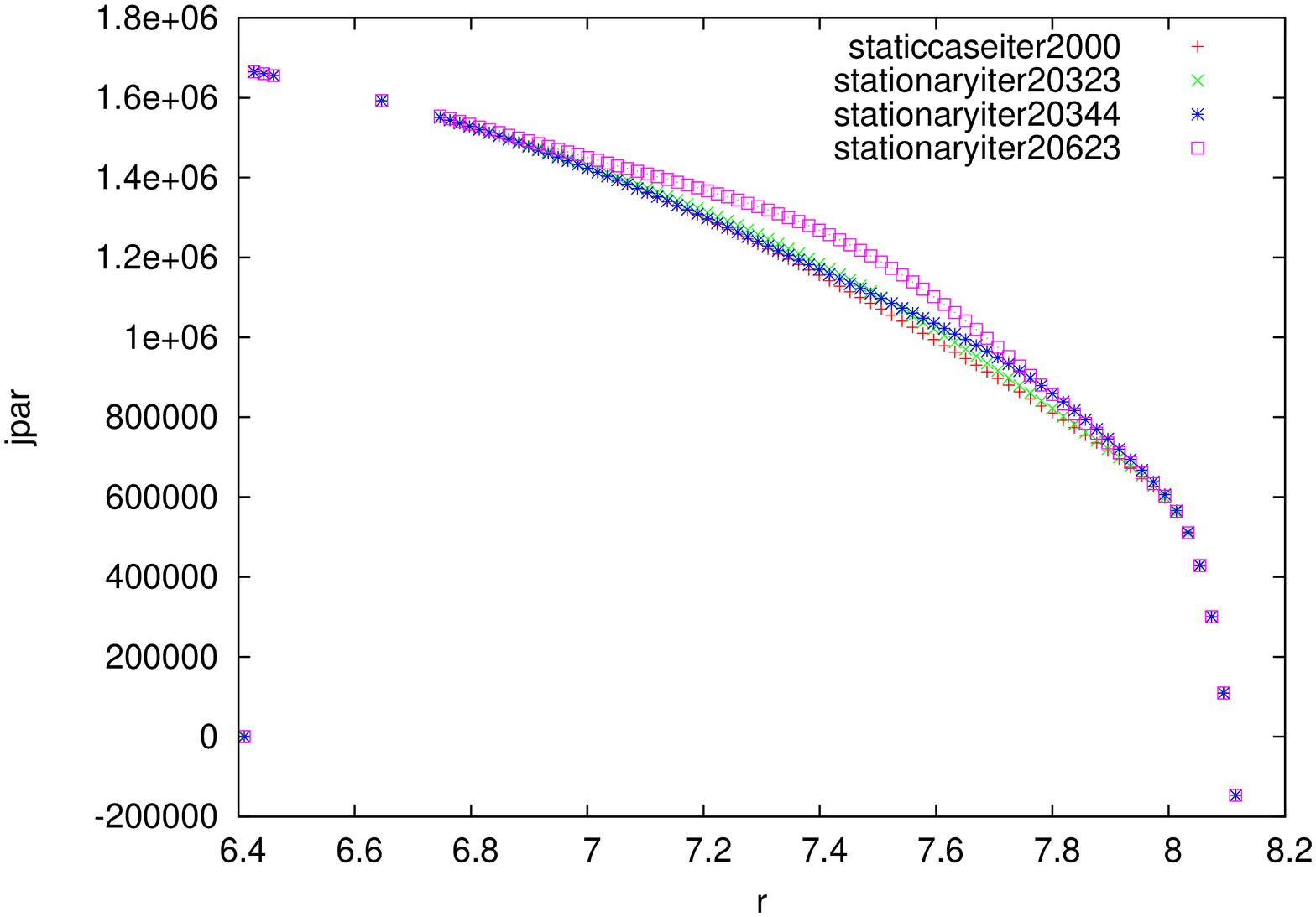}
\caption{Plots of the parallel to the magnetic field current density versus the 
radial distance from the axis of symmetry on the midplane $z=0$ for various  
Mach number profiles peaked on-axis.}
\label{fig:jpar1}
\end{center}
\end{figure}

\begin{figure}[ht!]
\begin{center}
\psfrag{jpar}{$J_\|(A/m^2)$}
\psfrag{r}{$R(m)$}
\psfrag{staticcaseiter2000 }[c][c]{Static Case, $M_0=0$\hspace{0.4cm}}
\psfrag{stationaryiter30224 }[c][c]{$M_0=0.02$, $m=2$, 
$n=4$\hspace{1cm}}
\psfrag{stationaryiter30552 }[c][c]{$M_0=0.05$, $m=5$, 
$n=2$\hspace{1cm}}
\includegraphics[scale=0.45, angle=270]{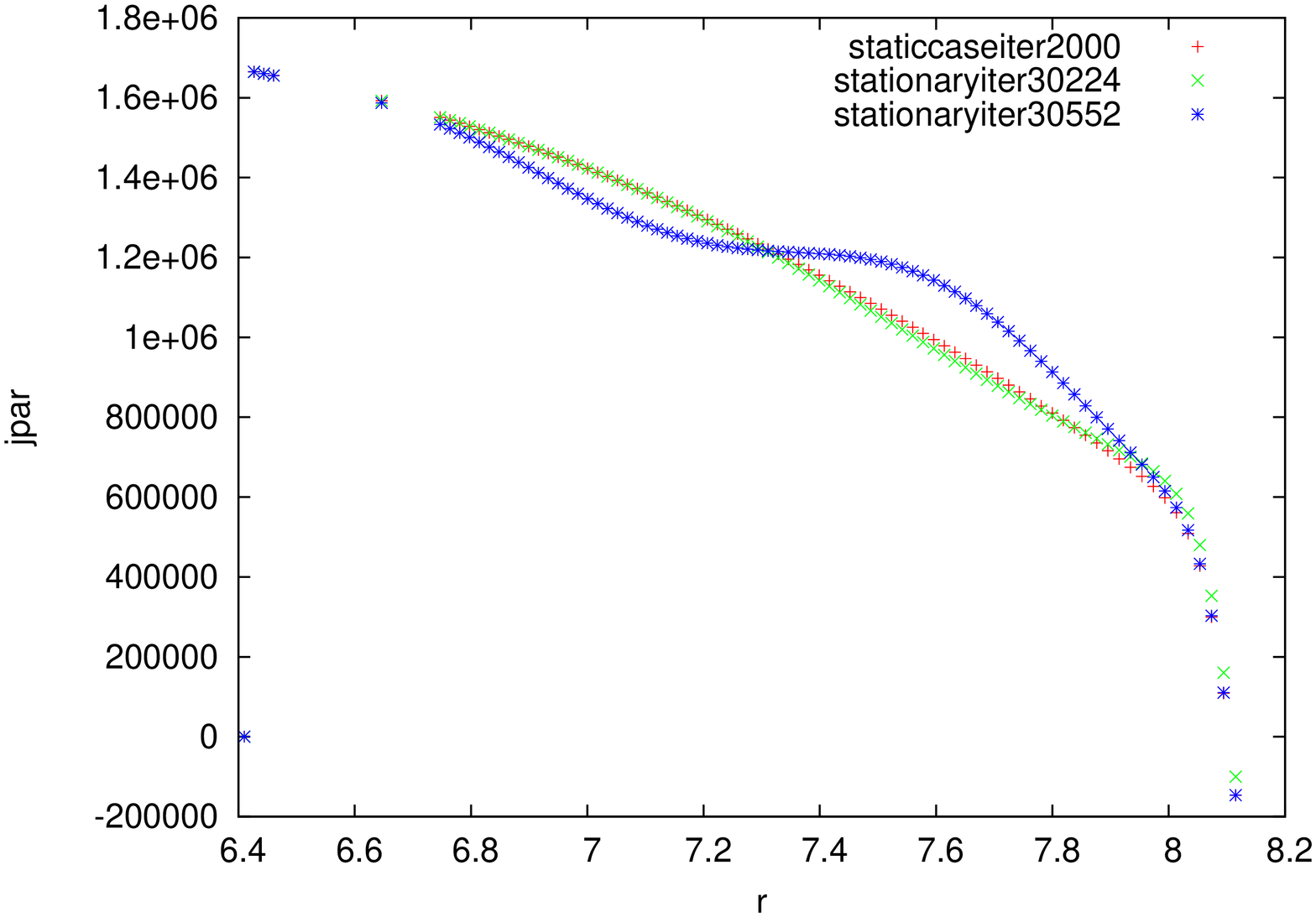}
\caption{Plots of the parallel to the magnetic field current density versus the 
radial distance from the axis of symmetry on the midplane $z=0$ for various 
Mach 
number profiles peaked off-axis.}
\label{fig:jpar2}
\end{center}
\end{figure}

Paying attention to the current density, one can conclude that the off-axis 
 rotation affects the profiles more than the on-axis as can be seen in Figs.  
\ref{fig:jphi1}- \ref{fig:jpar2}. This implies  that the  rotation shear is 
more 
important for the 
shaping of the plasma current density than the rotation itself. Therefore even 
for small values of the Mach number the impact of rotation shear on the current 
density can be 
significant. Comparison of Figs. \ref{fig:jphi1}-\ref{fig:jpar1} and 
\ref{fig:jphi2}-\ref{fig:jpar2} shows that for each individual Mach number 
profile the impact of parallel rotation on the parallel current density is 
stronger than on the toroidal component of the current density. This indicates 
that there is a correlation of the direction of the rotation and the 
corresponding direction of the current density. Experimentally, the co-current 
direction of rotation results in better quality ITBs for the electron 
temperature \cite{oyama2007ImpperlonELMH-mplainttrabarJT-}. Analysing the 
current density in two components, a toroidal and a poloidal one, 
 we can conclude that the plasma rotation affects more the poloidal component 
of 
the current density than the toroidal one. Taking into account that the plasma 
rotation is parallel to the magnetic field and the toroidal component of 
$\vec{B}$ is dominant over the poloidal one we conclude that also the toroidal 
component of the rotation is the dominant over the poloidal one and in turn 
affects the poloidal current density more than the toroidal current density. 

%-----------------------------
\section{Stability consideration}

We now consider the important issue of  linear stability for the equilibria 
constructed 
with respect to small  magnetohydrodynamic perturbations by applying the  
sufficient condition of 
\cite{throumoulopoulos2007sufconlinstamagequfiealiincflo_2, 
vladimirov1998thrstastemagfloideflu}. This condition 
states that a general steady state of a plasma of constant density and 
incompressible flow parallel to $\vec{ B}$ is linearly stable to small 
three-dimensional perturbations if the flow is sub-Alfv\'enic ($M^2<1$) and 
$A\geq 0$, where $A$ is given below by (\ref{Alpha})-(\ref{A4}). Note here that 
if the density is uniform at equilibrium it remains so at the perturbed state 
because of incompressibility.
For axisymmetric equilibria in the $\psi$-space $A$ assumes the form
\begin{align}
A=A_1+A_2+A_3+A_4 \label{Alpha} \\
A_1=-(1-M^2)\mu_0^2(\vec{J}\times \vec{\nabla} \psi)^2 \label{A1} \\
A_2= (1-M^2)\mu_0(\vec{J}\times \vec{\nabla} \psi)\cdot (\vec{\nabla} \psi\cdot 
\vec{\nabla})\vec{B} \label{A2} 
\\
A_3=-\frac{1}{2}\frac{dM^2}{d\psi}|\vec{\nabla} \psi|^2\left(\vec{\nabla} 
\psi\cdot \frac{\vec{\nabla} B^2}{2}\right) \label{A3} \\
A_4=-\frac{1}{2}\frac{dM^2}{d\psi}|\vec{\nabla} \psi|^{4}g \label{A4}
\end{align}
$$g= \frac{1}{1-M^2}\left(\frac{dP_s}{d\psi}-\frac{dM^2}{d\psi}\frac{B^2}{2 
\mu_0}
\right)$$
with $\vec{B}$ and $\vec{J}$ given by Eqs. (\ref{eq:9}) and (\ref{eq:10}).
The quantity $A_1$ being always negative consists a destabilizing
contribution  potentially related to current driven modes. The
other terms can be either stabilizing or destabilizing.
Specifically,  the term $A_2$ relates to the current density
and the variation of the magnetic field perpendicular to the
magnetic surfaces.   $A_3$ and $A_4$ are mostly flow terms
because they vanish in the absence of flow. $A_3$ additionally
depends on the  variation of the magnitude of the magnetic
field perpendicular to the magnetic surfaces while $A_4$
depends on the pressure gradient through the quantity $g$.
The flows satisfying (\ref{4}) are inherently sub-Alfv\'enic  because of the 
transformation (\ref{3}). Derivation of the condition is provided in 
\cite{throumoulopoulos2007sufconlinstamagequfiealiincflo_2}. Here it is just 
noted that it comes from the global requirement of non-negativeness of an 
integrand unlike  usual local requirements associated  with stability criteria 
for static equilibria, e.g. the Mercier criterion for localized interchanged 
modes. 

The condition was applied with the aid of Mathematica to equilibrium  outputs 
from HELENA which were interpolated by $5^{th}$ order splines assuring the 
necessary continuity of derivatives up to second order. Due to the fact that 
spline interpolation in Mathematica works only for structured grid an 
additional 
step was necessary. The appropriate output of HELENA was initially parsed to 
MATLAB which mapped it to a structured grid by means of Delaunay 
triangulation. The procedure was benchmarked against the    Solov\'ev solution 
showing accuracy up to order $10^{-10}$. The results show that for 17 flow 
profiles examined, including the static one, the condition is not satisfied 
within the plasma region. Evaluating the impact of plasma rotation on  the  
midplane $z=0$ at the points where the Mach number  exhibits its maximum and at 
the points where $|dM/d\psi|$ is maximum we concluded  that the rotation 
(through the terms $A_3$ and $A_4$) can be spatially in part destabilizing and 
in part destabilizing. 
\begin{figure}[ht!]
\begin{center}
\psfrag{A1}{$A_1$}
\psfrag{A2}{$A_2$}
\psfrag{A3}{$A_3$}
\psfrag{A4}{$A_4$}
\psfrag{R}{$R(m)$}
%\hspace{-2cm}
\centerline{\includegraphics[scale=1.1]{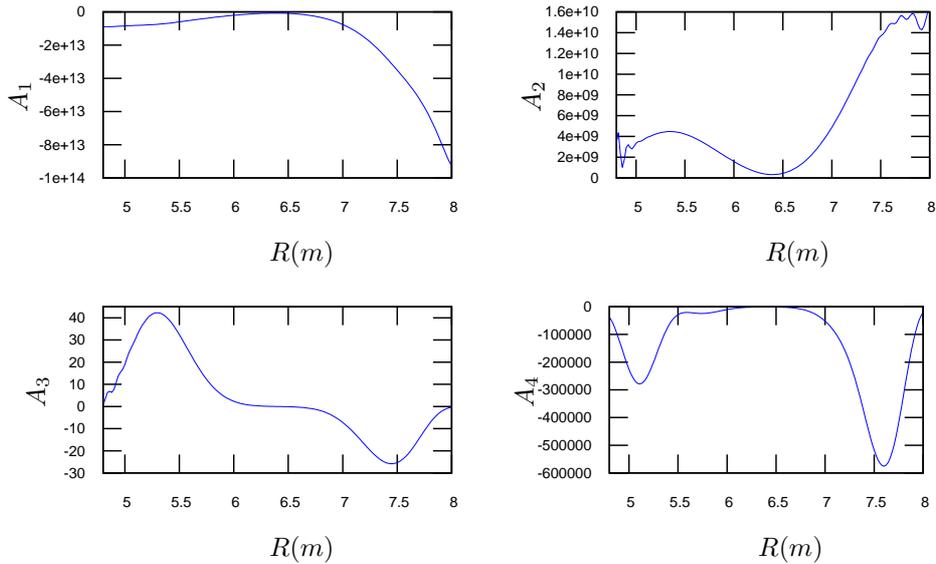}}
\caption{Plots of the quantities $A_1-A_4$ for a typical equilibrium with 
peaked 
on-axis 
Mach number (Eq. (\ref{prof1})) with  $M_0=0.06$, $n=2$, $m=3$.}
\label{fig:stab2}
\end{center}
\end{figure}
Also, this  can be seen in Fig. \ref{fig:stab2} where the individual terms 
consisting $A$ are plotted on the mid-plane $z=0$. The term $A_2$ related the 
magnetic shear is stabilizing but can not overcome the destabilizing term 
$A_1$. Also, the flow term $A_4$ has a weaker destabilizing contribution and 
$A_3$ is stabilizing in the inner region close to the axis of symmetry and 
destabilizing in the outer region.  It should be clarified here that the term 
``stabilizing" (``destabilizing") has the meaning of a non-negative (negative) 
contribution to $A$. However, since the condition is sufficient,  $A<0$ does 
not 
imply instability but  indecisiveness.  

In previous studies we made a similar stability study on the basis of analytic 
or quasianalytic solutions of the GGS equation (\ref{4}) for parallel rotation 
in plane \cite{throumoulopoulos2009Magcateyestaeffplaflo_2}, translational 
symmetric \cite{kuiroukidis2014Twononcylequrevmagshesheflo}  and axisymmetric 
toroidal \cite{kuiroukidis2015} geometries. In these studies it  turned out 
that 
the condition $A\geq 0$ is satisfied in a major part of the plasma. 
Stabilization is caused mainly  by the term $A_2$ in conjunction with 
equilibrium non-linearity with a weaker stabilizing contribution from the 
rotation and its shear.  Also, certain results indicate that the equilibrium 
non-linearity activates flow stabilization. In the present case, the input 
profiles are linear and that may explain the destabilizing results. In 
other studies  \cite{kuiroukidis2014ananontorequflo,Kaltsas2015} it turned out 
that the flow terms $A_3$ or $A_4$ are destabilizing. The above mentioned  
results and those of the present study clearly indicate that the impact of 
rotation and its shear on stability is far from universal; depending on the 
particular equilibria and the shape of the rotation profile the rotation can be 
either stabilizing or destabilizing.

%----------------------------
\section{Conclusions}

We extended the HELENA  fixed boundary equilibrium solver to equilibria with 
incompressible plasma rotation parallel to the magnetic field. The pertinent  
GGS equation by means of an integral transformation is put in a form  identical 
to the usual static GS equation. This transformation which maps the poloidal 
magnetic flux function $\psi$  to another flux function $u$ leaves the shape of 
the magnetic flux surfaces intact  just relabelling them. Via this and the 
inverse  transformation the calculated by the solver quantities in  the $u$ 
space, are mapped to the $\psi$-space. The code was run in the EFDA 
ITM-TF infrastructure with input shot 35441 developed for ITER simulations. A 
simple workflow in the Kepler environment was used for reading the input CPO of 
the shot and storing the results locally in the user's database.

On the basis of the equilibria constructed by the extended code we examined the 
impact of rotation and its shear on the pressure, toroidal current density and 
toroidal beta. As expected  the pressure is reduced  by the 
rotation. For localised flow the shape of the pressure profile 
resembles that observed in advanced confinement regimes. Specifically, owing 
to the plasma rotation, a rapid decrease in the pressure values accompanied by 
a flat region is evident around the maximum of the Mach number 
profile when this profile  is peaked off-axis. The effect of  peaked on-axis 
plasma rotation on the pressure profile  is weaker  due to the larger relative 
values of the  pressure  in the core region. The effect  of rotation  on the 
current density is  similar to that on the pressure with the additional remark 
that the impact of the parallel to the magnetic field rotation is 
stronger on the poloidal current density than that on the toroidal one. This 
result suggests that there is correlation between the direction of the rotation 
components and the corresponding components of the equilibrium current 
density. 

Furthermore, we examined the linear stability of the equilibria constructed by 
means of a sufficient condition. The results were inconclusive because the 
condition was satisfied nowhere in the plasma  for a variety of the  Mach 
number 
profiles examined. A detailed evaluation of the individual terms consisting 
the pertinent stability quantity $A$ (Eqs. (\ref{Alpha})-(\ref{A4})) showed 
that 
the variation of the magnetic field perpendicular to the magnetic surfaces, 
related to the magnetic shear, is stabilizing but, for the equilibria 
considered, can not overcome a destabilizing contribution potentially related 
to 
current driven modes. Also, depending on the spatial position a weaker impact 
of 
the flow can be either stabilizing or destabilizing in consistence with 
previous 
stability studies.

It is interesting to extend further the code in two respects: First, for non 
parallel incompressible flow associated with electric fields which play a role 
in the transitions to improved confinement regimes. This can be done on the 
basis of Eq. (\ref{4}) by including the $R^4$-term with the differential part 
of 
the equation remaining unaffected. Second, in the presence of pressure 
anisotropy which for compressible flows has already been considered in the FLOW 
code. 

\section{Aknowledgements}
 Part of this work was conducted
during visits of two of the authors (G.P. and G.N.T.) to the Max-Planck-Institut
f\"{u}r Plasmaphysik, Garching. The hospitality of that Institute
is greatly appreciated.

This work has been carried out within the framework of the EUROfusion
Consortium and has received funding from (a) the National Programme for the
Controlled Thermonuclear Fusion, Hellenic Republic, (b) Euratom research and
training programme 2014-2018 under grant agreement no. 633053. The views
and opinions expressed herein do not necessarily reflect those of the European
Commission.

\bibliographystyle{unsrtnat}
\bibliography{helena_par}

\begin{thebibliography}{41}
\providecommand{\natexlab}[1]{#1}
\providecommand{\url}[1]{\texttt{#1}}
\expandafter\ifx\csname urlstyle\endcsname\relax
  \providecommand{\doi}[1]{doi: #1}\else
  \providecommand{\doi}{doi: \begingroup \urlstyle{rm}\Url}\fi

\bibitem[Falchetto et~al.()Falchetto, Coster, Coelho, Scott, Figini, Kalupin,
  Nardon, Nowak, Alves, Artaud, Basiuk, Bizarro, Boulbe, Dinklage, Farina,
  Faugeras, Ferreira, Figueiredo, Huynh, Imbeaux, Ivanova-Stanik, Jonsson,
  Klingshirn, Konz, Kus, Marushchenko, Pereverzev, Owsiak, Poli, Peysson,
  Reimer, Signoret, Sauter, Stankiewicz, Strand, Voitsekhovitch, Westerhof,
  Zok, Zwingmann, Contributors, the Team, and {JET-EFDA
  Contributors}]{falchetto2014EurIntTokModITMeffachfirphyres}
GL~Falchetto, D~Coster, R~Coelho, BD~Scott, L~Figini, D~Kalupin, E~Nardon,
  S~Nowak, LL~Alves, JF~Artaud, V~Basiuk, Jo\~{a}o~PS Bizarro, C~Boulbe,
  A~Dinklage, D~Farina, B~Faugeras, J~Ferreira, A~Figueiredo, Ph~Huynh,
  F~Imbeaux, I~Ivanova-Stanik, T~Jonsson, H-J Klingshirn, C~Konz, A~Kus,
  NB~Marushchenko, G~Pereverzev, M~Owsiak, E~Poli, Y~Peysson, R~Reimer,
  J~Signoret, O~Sauter, R~Stankiewicz, P~Strand, I~Voitsekhovitch, E~Westerhof,
  T~Zok, W~Zwingmann, ITM-TF Contributors, ASDEX~Upgrade the Team, and
  {JET-EFDA Contributors}.
\newblock {The European Integrated Tokamak Modelling (ITM) effort: achievements
  and first physics results}.
\newblock \emph{Nuclear Fusion}, \penalty0 (4):\penalty0 043018.
\newblock \doi{10.1088/0029-5515/54/4/043018}.

\bibitem[G\"{u}nter et~al.()G\"{u}nter, Wolf, Leuterer, Gruber, Kaufmann,
  Lackner, Maraschek, {Mc Carthy}, Meister, Peeters,
  et~al.]{gunter2000SimatthigeleiontemdisinttrabarASDupg}
S~G\"{u}nter, RC~Wolf, F~Leuterer, O~Gruber, M~Kaufmann, K~Lackner,
  M~Maraschek, PJ~{Mc Carthy}, H~Meister, A~Peeters, et~al.
\newblock {Simultaneous attainment of high electron and ion temperatures in
  discharges with internal transport barriers in ASDEX upgrade}.
\newblock \emph{Physical Review Letters}, \penalty0 (14):\penalty0 3097.

\bibitem[Romanelli et~al.()Romanelli, Lax{\aa}back,
  et~al.]{romanelli2011OveJETres}
F~Romanelli, M~Lax{\aa}back, et~al.
\newblock {Overview of JET results}.
\newblock \emph{Nuclear Fusion}, \penalty0 (9):\penalty0 094008.

\bibitem[Itoh and Itoh()]{itoh1996rolelefiecon}
K~Itoh and SI~Itoh.
\newblock {The role of the electric field in confinement}.
\newblock \emph{Plasma Physics and Controlled Fusion}, \penalty0 (1):\penalty0
  1--49.

\bibitem[{Challis}(2004)]{challis2004useinttrabartokpla_2}
CD~{Challis}.
\newblock {The use of internal transport barriers in tokamak plasmas}.
\newblock \emph{Plasma Physics and Controlled Fusion}, 46:\penalty0 B23--B40,
  December 2004.
\newblock \doi{10.1088/0741-3335/46/12B/003}.

\bibitem[Sips et~al.()Sips, Tardini, Forest, Gruber, {Mc Carthy}, Gude, Horton,
  Igochine, Kardaun, Maggi, et~al.]{sips2007perimpH-mASDUpgprotoITE}
ACC Sips, G~Tardini, CB~Forest, O~Gruber, PJ~{Mc Carthy}, A~Gude, LD~Horton,
  V~Igochine, O~Kardaun, CF~Maggi, et~al.
\newblock {The performance of improved H-modes at ASDEX Upgrade and projection
  to ITER}.
\newblock \emph{Nuclear Fusion}, \penalty0 (11):\penalty0 1485.

\bibitem[Solomon et~al.()Solomon, Burrell, Budny, Groebner, Kinsey, Kramer,
  Luce, Makowski, Mikkelsen, Nazikian, et~al.]{solomon2007Momconlowtor}
WM~Solomon, KH~Burrell, R~Budny, RJ~Groebner, JE~Kinsey, GJ~Kramer, TC~Luce,
  MA~Makowski, D~Mikkelsen, R~Nazikian, et~al.
\newblock {Momentum confinement at low torque}.
\newblock \emph{Plasma Physics and Controlled Fusion}, \penalty0
  (12B):\penalty0 B313.

\bibitem[Eriksson et~al.()Eriksson, Righi, and
  Zastrow]{eriksson1997TorrotICRH-mJET}
L-G Eriksson, E~Righi, and K-D Zastrow.
\newblock {Toroidal rotation in ICRF-heated H-modes on JET}.
\newblock \emph{Plasma Physics and Controlled Fusion}, \penalty0 (1):\penalty0
  27--42.

\bibitem[{Morozov} and {Solov{\rq}ev}()]{morozov1980Steplaflomagfie}
AI~{Morozov} and LS~{Solov{\rq}ev}.
\newblock {Steady-state plasma flow in a magnetic field}.
\newblock In \emph{{Reviews of Plasma Physics}}, pages 1--103. Springer.

\bibitem[{Hameiri}(1983)]{hameiri1983equstarotpla}
E~{Hameiri}.
\newblock {The equilibrium and stability of rotating plasmas}.
\newblock \emph{Physics of Fluids}, 26:\penalty0 230--237, January 1983.
\newblock \doi{10.1063/1.864012}.

\bibitem[Semenzato et~al.()Semenzato, Gruber, and
  Zehrfeld]{semenzato1984ComsymideMHDfloequ}
S~Semenzato, R~Gruber, and HP~Zehrfeld.
\newblock {Computation of symmetric ideal MHD flow equilibria}.
\newblock \emph{Computer Physics Reports}, \penalty0 (7):\penalty0 389--425.

\bibitem[Strumberger et~al.()Strumberger, G\"{u}nter, Merkel, Riondato,
  Schwarz, Tichmann, and
  Zehrfeld]{strumberger2005NumMHDstastutorrotvisreswalcurhol_2}
E~Strumberger, S~G\"{u}nter, P~Merkel, S~Riondato, E~Schwarz, C~Tichmann, and
  HP~Zehrfeld.
\newblock {Numerical MHD stability studies: toroidal rotation, viscosity,
  resistive walls and current holes}.
\newblock \emph{Nuclear Fusion}, \penalty0 (9):\penalty0 1156.

\bibitem[Beli\"{e}n et~al.()Beli\"{e}n, Botchev, Goedbloed, van~der Holst, and
  Keppens]{belien2002FINAxiMHDEquFlo}
AJC Beli\"{e}n, MA~Botchev, JP~Goedbloed, B~van~der Holst, and R~Keppens.
\newblock {FINESSE: Axisymmetric MHD Equilibria with Flow}.
\newblock \emph{J. Comput. Phys.}, \penalty0 (1):\penalty0 91--117, October .
\newblock ISSN 0021-9991.
\newblock \doi{10.1006/jcph.2002.7153}.

\bibitem[Guazzotto et~al.()Guazzotto, Betti, Manickam, and
  Kaye]{guazzotto2004Numstutokequarbflo}
L~Guazzotto, R~Betti, J~Manickam, and S~Kaye.
\newblock {Numerical study of tokamak equilibria with arbitrary flow}.
\newblock \emph{Physics of Plasmas}, \penalty0 (2):\penalty0 604--614, February
  .
\newblock \doi{10.1063/1.1637918}.

\bibitem[Courant and Hilbert()]{courant1966Metmatphy}
R~Courant and D~Hilbert.
\newblock \emph{{Methods of mathematical physics vol. 2}}.
\newblock CUP Archive.

\bibitem[{Budny} et~al.(1995){Budny}, {Bell}, {Janos}, {Jassby}, {Johnson},
  {Mansfield}, {McCune}, {Redi}, {Schivell}, {Taylor}, {Terpstra},
  {Zarnstorff}, and {Zweben}]{budny1995SimalpparTFTDTsuphigfuspow}
RV~{Budny}, MG~{Bell}, AC~{Janos}, DL~{Jassby}, LC~{Johnson}, DK~{Mansfield},
  DC~{McCune}, MH~{Redi}, JF~{Schivell}, G~{Taylor}, TB~{Terpstra},
  MC~{Zarnstorff}, and SJ~{Zweben}.
\newblock {Simulations of alpha parameters in a TFTR DT supershot with high
  fusion power}.
\newblock \emph{Nuclear Fusion}, 35:\penalty0 1497--1508, 1995.
\newblock \doi{10.1088/0029-5515/35/12/I10}.

\bibitem[Friedlander and Vishik()]{friedlander1995stainscrimag}
S~Friedlander and MM~Vishik.
\newblock {On stability and instability criteria for magnetohydrodynamics}.
\newblock \emph{Chaos: An Interdisciplinary Journal of Nonlinear Science},
  \penalty0 (2):\penalty0 416--423.

\bibitem[Morrison et~al.()Morrison, Tassi, and
  Tronko]{morrison2013Stacomredmagequmagins}
PJ~Morrison, E~Tassi, and N~Tronko.
\newblock {Stability of compressible reduced magnetohydrodynamic
  equilibria---Analogy with magnetorotational instability}.
\newblock \emph{Physics of Plasmas}, \penalty0 (4):\penalty0 042109.

\bibitem[Garofalo et~al.({\natexlab{a}})Garofalo, Turnbull, Austin, Bialek,
  Chu, Comer, Fredrickson, Groebner, {La Haye}, Lao,
  et~al.]{garofalo1999Dirobsreswalmodtokitsintplarot}
AM~Garofalo, AD~Turnbull, ME~Austin, J~Bialek, MS~Chu, KJ~Comer,
  ED~Fredrickson, RJ~Groebner, RJ~{La Haye}, LL~Lao, et~al.
\newblock {Direct observation of the resistive wall mode in a tokamak and its
  interaction with plasma rotation}.
\newblock \emph{Physical Review Letters}, \penalty0 (19):\penalty0 3811,
  {\natexlab{a}}.

\bibitem[Garofalo et~al.({\natexlab{b}})Garofalo, Jackson, {La Haye},
  Okabayashi, Reimerdes, Strait, Ferron, Groebner, In, Lanctot,
  et~al.]{garofalo2007StaconreswalmodhigbetlowrotDIIpla}
AM~Garofalo, GL~Jackson, RJ~{La Haye}, M~Okabayashi, H~Reimerdes, EJ~Strait,
  JR~Ferron, RJ~Groebner, Y~In, MJ~Lanctot, et~al.
\newblock {Stability and control of resistive wall modes in high beta, low
  rotation DIII-D plasmas}.
\newblock \emph{Nuclear Fusion}, \penalty0 (9):\penalty0 1121, {\natexlab{b}}.

\bibitem[Chu and Okabayashi()]{chu2010Staextkinreswalmod}
MS~Chu and M~Okabayashi.
\newblock {Stabilization of the external kink and the resistive wall mode}.
\newblock \emph{Plasma Physics and Controlled Fusion}, \penalty0 (12):\penalty0
  123001.

\bibitem[Igochine()]{igochine2012Phyreswalmod}
V~Igochine.
\newblock {Physics of resistive wall modes}.
\newblock \emph{Nuclear Fusion}, \penalty0 (7):\penalty0 074010.
\newblock \doi{10.1088/0029-5515/52/7/074010}.

\bibitem[Politzer et~al.()Politzer, Petty, Jayakumar, Luce, Wade, DeBoo,
  Ferron, Gohil, Holcomb, Hyatt,
  et~al.]{politzer2008InftorrottrastahybsceplaDII}
PA~Politzer, CC~Petty, RJ~Jayakumar, TC~Luce, MR~Wade, JC~DeBoo, JR~Ferron,
  P~Gohil, CT~Holcomb, AW~Hyatt, et~al.
\newblock {Influence of toroidal rotation on transport and stability in hybrid
  scenario plasmas in DIII-D}.
\newblock \emph{Nuclear Fusion}, \penalty0 (7):\penalty0 075001.

\bibitem[{La Haye} et~al.(){La Haye}, Brennan, Buttery, and
  Gerhardt]{lahaye2010Islstreffplafloteasta}
RJ~{La Haye}, DP~Brennan, RJ~Buttery, and SP~Gerhardt.
\newblock {Islands in the stream: The effect of plasma flow on tearing
  stability}.
\newblock \emph{Physics of Plasmas}, \penalty0 (5):\penalty0 056110.

\bibitem[Konz et~al.(2011)Konz, Zwingmann, Osmanlic, Guillerminet, Imbeaux,
  Huynh, Plociennik, Owsiak, Zok, and
  Dunne]{konz2011FirphyappIntTokModITMtootoMHDstaanaexpdatITEsce}
C~Konz, W~Zwingmann, F~Osmanlic, B~Guillerminet, F~Imbeaux, P~Huynh,
  M~Plociennik, M~Owsiak, T~Zok, and M~Dunne.
\newblock {First physics applications of the Integrated Tokamak Modelling
  (ITM-TF) tools to the MHD stability analysis of experimental data and ITER
  scenarios}.
\newblock \emph{EPS}, page~O2, 2011.
\newblock URL \url{http://ocs.ciemat.es/eps2011pap/pdf/O2.103.pdf}.

\bibitem[Throumoulopoulos and
  Tasso()]{throumoulopoulos2007sufconlinstamagequfiealiincflo_2}
GN~Throumoulopoulos and H~Tasso.
\newblock {A sufficient condition for the linear stability of
  magnetohydrodynamic equilibria with field aligned incompressible flows}.
\newblock \emph{Physics of Plasmas}, \penalty0 (12):\penalty0 122104.

\bibitem[Tasso and Throumoulopoulos()]{tasso1998Axiidemagequincflo_2}
H~Tasso and GN~Throumoulopoulos.
\newblock {Axisymmetric ideal magnetohydrodynamic equilibria with
  incompressible flows}.
\newblock \emph{Physics of Plasmas}, \penalty0 (6):\penalty0 2378--2383.

\bibitem[Poulipoulis et~al.()Poulipoulis, Throumoulopoulos, and
  Tasso]{poulipoulis2005Torflochamagtopequeig}
G~Poulipoulis, GN~Throumoulopoulos, and H~Tasso.
\newblock {Toroidal flow-caused change in magnetic topology of equilibrium
  eigenstates}.
\newblock \emph{Physics of Plasmas}, \penalty0 (4):\penalty0 042112.
\newblock \doi{10.1063/1.1867497}.

\bibitem[{Throumoulopoulos} et~al.(2008){Throumoulopoulos}, {Tasso}, and
  {Poulipoulis}]{throumoulopoulos2008Sidaxiequincflo}
GN~{Throumoulopoulos}, H~{Tasso}, and G~{Poulipoulis}.
\newblock {Side-conditioned axisymmetric equilibria with incompressible flows}.
\newblock \emph{Journal of Plasma Physics}, 74:\penalty0 327--344, June 2008.
\newblock \doi{10.1017/S0022377807006769}.

\bibitem[{Throumoulopoulos} and
  {Tasso}(2003)]{throumoulopoulos2003axiresmagequflofrePfidif}
GN~{Throumoulopoulos} and H~{Tasso}.
\newblock {On axisymmetric resistive magnetohydrodynamic equilibria with flow
  free of Pfirsch-Schl\"{u}ter diffusion}.
\newblock \emph{Physics of Plasmas}, 10:\penalty0 2382--2388, June 2003.
\newblock \doi{10.1063/1.1571542}.

\bibitem[Huysmans et~al.()Huysmans, Goedbloed, and
  Kerner]{huysmans1991IsobicHerelesolGraequ}
GTA Huysmans, JP~Goedbloed, and W~Kerner.
\newblock {Isoparametric bicubic Hermite elements for solution of the
  Grad-Shafranov equation}.
\newblock \emph{International Journal of Modern Physics C}, \penalty0
  (01):\penalty0 371--376.

\bibitem[{Konz} and {Zille}(2007)]{KonzHELENA}
C~{Konz} and R~{Zille}.
\newblock \emph{{Manual of HELENA Fixed Boundary Equilibrium Solver}}.
\newblock Max-Planck Institute for Plasma Physics, 2007.

\bibitem[Simintzis et~al.()Simintzis, Throumoulopoulos, Pantis, and
  Tasso]{simintzis2001Anamagequmagconplasheflo}
Ch~Simintzis, GN~Throumoulopoulos, G~Pantis, and H~Tasso.
\newblock {Analytic magnetohydrodynamic equilibria of a magnetically confined
  plasma with sheared flows}.
\newblock \emph{Physics of Plasmas}, \penalty0 (6):\penalty0 2641--2648.

\bibitem[de~Vries et~al.()de~Vries, Hua, McDonald, Giroud, Janvier, Johnson,
  Tala, Zastrow, and Contributors]{devries2008ScarotmomconJETpla}
Peter~C de~Vries, M-D Hua, DC~McDonald, C~Giroud, M~Janvier, MF~Johnson,
  T~Tala, K-D Zastrow, and JET~EFDA Contributors.
\newblock {Scaling of rotation and momentum confinement in JET plasmas}.
\newblock \emph{Nuclear Fusion}, \penalty0 (6):\penalty0 065006.

\bibitem[Oyama et~al.()Oyama, Isayama, Suzuki, Koide, Takenaga, Ide, Nakano,
  Asakura, Kubo, Takechi, et~al.]{oyama2007ImpperlonELMH-mplainttrabarJT-}
N~Oyama, A~Isayama, T~Suzuki, Y~Koide, H~Takenaga, S~Ide, T~Nakano, N~Asakura,
  H~Kubo, M~Takechi, et~al.
\newblock {Improved performance in long-pulse ELMy H-mode plasmas with internal
  transport barrier in JT-60U}.
\newblock \emph{Nuclear Fusion}, \penalty0 (7):\penalty0 689.

\bibitem[Vladimirov and Ilin()]{vladimirov1998thrstastemagfloideflu}
VA~Vladimirov and KI~Ilin.
\newblock {The three-dimensional stability of steady magnetohydrodynamic flows
  of an ideal fluid}.
\newblock \emph{Physics of Plasmas}, \penalty0 (12):\penalty0 4199--4204.

\bibitem[{Throumoulopoulos} et~al.(2009){Throumoulopoulos}, {Tasso}, and
  {Poulipoulis}]{throumoulopoulos2009Magcateyestaeffplaflo_2}
GN~{Throumoulopoulos}, H~{Tasso}, and G~{Poulipoulis}.
\newblock {Magnetohydrodynamic 'cat eyes' and stabilizing effects of plasma
  flow}.
\newblock \emph{Journal of Physics A Mathematical General}, 42:\penalty0
  335501, August 2009.
\newblock \doi{10.1088/1751-8113/42/33/335501}.

\bibitem[{Kuiroukidis} and
  {Throumoulopoulos}(2014)]{kuiroukidis2014Twononcylequrevmagshesheflo}
Ap~{Kuiroukidis} and GN~{Throumoulopoulos}.
\newblock {Two-dimensional nonlinear cylindrical equilibria with reversed
  magnetic shear and sheared flow}.
\newblock \emph{Journal of Plasma Physics}, 80:\penalty0 27--41, February 2014.
\newblock \doi{10.1017/S0022377813000883}.

\bibitem[{Kuiroukidis} and {Throumoulopoulos}({\natexlab{a}})]{kuiroukidis2015}
Ap~{Kuiroukidis} and GN~{Throumoulopoulos}.
\newblock { Toroidal equilibrium states with reversed magnetic shear and
  parallel flow in connection with the formation of Internal Transport
  Barriers}.
\newblock \emph{Journal of Plasma Physics}, \penalty0 (4):\penalty0 905810404,
  {\natexlab{a}}.

\bibitem[{Kuiroukidis} and
  {Throumoulopoulos}({\natexlab{b}})]{kuiroukidis2014ananontorequflo}
Ap~{Kuiroukidis} and GN~{Throumoulopoulos}.
\newblock {An analytic nonlinear toroidal equilibrium with flow}.
\newblock \emph{Plasma Physics and Controlled Fusion}, \penalty0 (7):\penalty0
  75003--75009, {\natexlab{b}}.
\newblock \doi{10.1088/0741-3335/56/7/075003}.

\bibitem[{Kaltsas} and {Throumoulopoulos}()]{Kaltsas2015}
DA~{Kaltsas} and GN~{Throumoulopoulos}.
\newblock {Generalized Solovev equilibrium with sheared flow of arbitrary
  direction and stability consideration}.
\newblock \emph{Physics of Plasmas}, page 084502.

\end{thebibliography}

\end{document}